\documentclass[preprint,aps,prc,tightenlines,nofootinbib]{revtex4}
\usepackage{amssymb}
\usepackage{bm}
\begin{document}
\input{psfig}
\input{epsf}
\def\Im{\mbox{\sl Im\ }}
\def\pd{\partial}
\def\oln{\overline}
\def\olft{\overleftarrow}
\def\ds{\displaystyle}
\def\bgreek#1{\mbox{\boldmath $#1$ \unboldmath}}
\def\sla#1{\slash \hspace{-2.5mm} #1}
\newcommand{\bra}{\langle}
\newcommand{\ket}{\rangle}
\newcommand{\vep}{\varepsilon}
\newcommand{\met}{{\mbox{\scriptsize met}}}
\newcommand{\lab}{{\mbox{\scriptsize lab}}}
\newcommand{\cm}{{\mbox{\scriptsize cm}}}
\newcommand{\mcal}{\mathcal}
\newcommand{\Del}{$\Delta$}
\newcommand{\g}{{\rm g}}
\long\def\Omit#1{}
\long\def\omit#1{\small #1}
\def\beq{\begin{equation}}
\def\eeq{\end{equation} }
\def\bea{\begin{eqnarray}}
\def\eea{\end{eqnarray}}
\def\eqref#1{Eq.~(\ref{eq:#1})}
\def\eqlab#1{\label{eq:#1}}
\def\figref#1{Fig.~\ref{fig:#1}}
\def\figlab#1{\label{fig:#1}}
\def\tabref#1{Table \ref{tab:#1}}
\def\tablab#1{\label{tab:#1}}
\def\secref#1{Section~\ref{sec:#1}}
\def\appref#1{Appendix~\ref{sec:#1}}
\def\seclab#1{\label{sec:#1}}
\def\VYP#1#2#3{{\bf #1}, #3 (#2)}  
\def\NP#1#2#3{Nucl.~Phys.~\VYP{#1}{#2}{#3}}
\def\NPA#1#2#3{Nucl.~Phys.~A~\VYP{#1}{#2}{#3}}
\def\NPB#1#2#3{Nucl.~Phys.~B~\VYP{#1}{#2}{#3}}
\def\PL#1#2#3{Phys.~Lett.~\VYP{#1}{#2}{#3}}
\def\PLB#1#2#3{Phys.~Lett.~B~\VYP{#1}{#2}{#3}}
\def\PR#1#2#3{Phys.~Rev.~\VYP{#1}{#2}{#3}}
\def\PRA#1#2#3{Phys.~Rev.~A~\VYP{#1}{#2}{#3}}
\def\PRC#1#2#3{Phys.~Rev.~C~\VYP{#1}{#2}{#3}}
\def\PRD#1#2#3{Phys.~Rev.~D~\VYP{#1}{#2}{#3}}
\def\PRL#1#2#3{Phys.~Rev.~Lett.~\VYP{#1}{#2}{#3}}
\def\FBS#1#2#3{Few-Body~Sys.~\VYP{#1}{#2}{#3}}
\def\AP#1#2#3{Ann.~of Phys.~\VYP{#1}{#2}{#3}}
\def\ZP#1#2#3{Z.\ Phys.\  \VYP{#1}{#2}{#3}}
\def\ZPA#1#2#3{Z.\ Phys.\ A\VYP{#1}{#2}{#3}}
\def\JPA#1#2#3{J.~Phys.~A\VYP{#1}{#2}{#3}}
\def\half{\mbox{\small{$\frac{1}{2}$}}}
\def\quarter{\mbox{\small{$\frac{1}{4}$}}}
\def\nn{\nonumber}
\newlength{\PicSize}
\newlength{\FormulaWidth}
\newlength{\DiagramWidth}
\newcommand{\vslash}[1]{#1 \hspace{-0.42 em} /}
\newcommand{\qslash}[1]{#1 \hspace{-0.46 em} /}
\def\her{\marginpar{$\Longleftarrow$}}
\def\bel{\marginpar{$\Downarrow$}}
\def\abo{\marginpar{$\Uparrow$}}



\title{Interpretation of the neutron quantum gravitational states in terms of
isospectral potentials} 

\author{S.~Kondratyuk}
\affiliation{Department of Physics and Astronomy, University of Manitoba,
Winnipeg, MB, Canada R3T 2N2}
\author{P.~G.~Blunden}
\affiliation{Department of Physics and Astronomy, University of Manitoba,
Winnipeg, MB, Canada R3T 2N2}

\date{\today}

\begin{abstract}
The recently observed quantum states of neutrons bound in a gravitational field are analyzed in the framework of one-parameter isospectral hamiltonians. Potentials isospectral to the usual Newton potential are explicitly constructed for the first time, then constrained using measured properties of the neutron gravitational states. The corresponding wave functions and the neutron fluxes are also calculated and analyzed in a simple model, including the ground state and the excited state contributions. 
The constructed isospectral potentials are discussed as 
candidates for a possible modification of Newton's law at a submillimetre scale.    
Our results indicate that significant deviations from the Newtonian gravity at submillimetre distances 
could be compatible with experiment.
\end{abstract}

\maketitle


\section{Introduction} \seclab{intro}

In recent years there has been a significant progress in
the study of gravity on a submillimetre scale. It has been precipitated by the first 
measurements~\cite{Nes02,Nes05} of the quantum levels of ultra-cold neutrons propagating under gravity
within a submillimetre-wide slit between a bottom mirror and a top absorber.
These results have profound implications as they could put experimental constraints on proposed extensions of the Standard Model (see e.~g.~\cite{Ark98,Bur04,Bui07} and references cited therein). 
 
The data on the quantum gravitational states of neutrons have been analyzed in several 
phenomenological models~\cite{Nes02,Nes05,Vor06,Wes07,Nes04,Bae07,Adh07} with the aim of exploring possible deviations from Newton's law at a submillimetre scale. Although the data can be explained in models with unmodified gravity and a more-or-less sophisticated description of the neutron guide 
system~\cite{Nes05,Vor06,Wes07,Adh07}, the current experimental uncertainties do leave some room for possible modifications of Newton's law at submillimetre 
distances~\cite{Nes04,Bae07}.    

In this paper we consider this problem from the point of view of quantum-mechanical isospectral hamiltonians. 
Construction methods and various applications of isospectral hamiltonians can be found in 
Refs.~\cite{Abr80,Cru54,Kha89} and in monographs~\cite{Bag01,Coo01}.
An important feature of isospectral hamiltonians is that despite having quite distinct potentials 
(which we will call ``isospectral potentials"), they all have identical energy spectra.  
This remarkable property lies at the core of our motivation to explore possible deviations from the Newton
law by studying potentials isospectral to the usual gravitational potential. 
In this way we ensure that 
any possible modification of the Newton potential preserves the quantum-mechanical energy spectrum and is
thus compatible with the measured properties of the gravitational levels of neutrons.
 
We will first construct a one-parameter family of quantum-mechanical hamiltonians isospectral to the hamiltonian with the usual Newtonian potential of a neutron in the gravitational field, and calculate the corresponding wave functions of the ground and excited states. 
We will use measured properties of the neutron
gravitational quantum states~\cite{Nes02,Nes05} as experimental constraints to select only those isospectral hamiltonians which are compatible with these experimental data. Specifically, the characteristic step-like dependence of the measured neutron flux on the width of the slit will be used to select the most likely isospectral potentials.
We will explore the possibility that these isospectral potentials could describe deviations from the Newtonian gravitational law at a submillimetre scale.       
(There are important qualifications which limit the conclusions drawn from a comparison of the present simplified model with the existing data;
this will be explained in the text, and we will suggest a modified experimental arrangement to more directly test our results.) 
The results of our model analysis indicate that the experimental data allow for appreciable modifications of Newton`s gravitational potential at heights of the order of tens of micrometres.

\section{One-parameter isospectral gravitational hamiltonians}\seclab{isospec}

We consider a neutron of mass $m$, at a height $z$ over the surface of a perfect 
mirror. Such a system is described by a one-dimensional hamiltonian 
\beq
H(z) = -{\hbar^2 \over {2m}} {d^2 \over {d z^2}} + V(z) \,,
\eqlab{ham_orig}
\eeq 
with the potential 
\beq
V(z \ge 0) = m g z, \;\;\; V(z < 0)=\infty \,, 
\eqlab{pot_orig}
\eeq
where $g$ is the usual  gravitational acceleration. 

The solution of this quantum-mechanical problem is 
well known~\cite{Flu74}. The neutron energy levels $E_n$ are proportional
to the zeros of the Airy function $Ai$~\cite{Abr72},
$\lambda_1 \approx 2.34, \, \lambda_2 \approx 4.09, \, \lambda_3 \approx 5.52\, \ldots$
($Ai(-\lambda_n)=0$):
\beq
E_n = \lambda_n \left[ m g^2 \hbar^2/2 \right]^{1/3}, \, n=1,2,3,\ldots \,,
\eqlab{energies}
\eeq
or numerically:  
$E_1 \approx 1.41 \, {\rm peV} \approx 2.25 \times 10^{-24} \, {\rm erg}, \, 
E_2 \approx 2.46 \, {\rm peV} \approx 3.92 \times 10^{-24} \, {\rm erg}, \, 
E_3 \approx 3.32 \, {\rm peV} \approx 5.29 \times 10^{-24} \, {\rm erg}, \ldots$
These energy levels can be also written with high precision in the WKB approximation; 
see~\cite{Flu74} for details.
The neutron wave functions are  
\beq
\psi_n(z \ge 0) = C_n Ai\left({z \over l} - \lambda_n\right)\,, \;\;\; \psi_n(z < 0) = 0 \,,
\eqlab{airy}
\eeq  
where $l = \left[ \hbar^2/(2 m^2 g) \right]^{1/3} \approx 5.87 \mu m$ is the characteristic length and the  
constants $C_n$ are determined from the normalization condition
$ \int_0^\infty \! dz \psi_n^2(z) = 1$. 
 
We are interested in a family of hamiltonians, parametrized by a real number $p$, 
\beq
H(z;p) = -{\hbar^2 \over {2m}} {\pd^2 \over {\pd z^2}} + V(z;p) \,,
\eqlab{iso_ham}
\eeq
which have the same energy spectrum $E_n$ as
the original hamiltonian $H(z)$. Many derivations of such 
isospectral hamiltonians have been developed in the past 
(see~\cite{Abr80,Cru54,Kha89} and references therein). One modern approach 
is usually formulated in the language
of supersymmetric quantum mechanics~\cite{Kha89,Bag01,Coo01}.    
Since the focus of this paper is the problem of neutron gravitational states, rather than isospectral hamiltonians in general, 
we will simply use the necessary results from~\cite{Coo01} without proof.

Given the original potential $V(z)$ and the normalized ground state wave function 
$\psi_1(z) \equiv \psi(z)$, the isospectral potential in \eqref{iso_ham} can be written as
\beq
V(z;p) = V(z) - {\hbar^2 \over m}{\pd^2 \over \pd z^2} 
\left[ {\rm ln} \left( {\mathcal{I}}(z) + p \right) \right] \,,
\eqlab{isopot}
\eeq
where  
\beq
{\mathcal{I}}(z) = \int\limits_0^z \! d z' \psi^2(z') \,.
\eqlab{int}
\eeq
Since $\psi(z)$ is normalized to unity, it follows 
from~\eqref{int} that ${\mathcal{I}}(z) \in [0,1]$ and therefore, to ensure that $V(z;p)$ 
in~\eqref{isopot} is finite, $p \notin [-1,0]$. 

Several members of the one-parameter family of potentials $V(z;p)$ isospectral to the linear Newtonian potential $m g z$ are shown in~\figref{isopot}, demonstrating the deformation of the potential with varying parameter $p$. (The figure will be further explained and discussed in the next section.) 
To our knowledge, this is the first time that these isospectral potentials have been explicitly calculated.   
Being isospectral to the exactly solvable quantum-mechanical system with the potential $m g z$, each of the potentials in~\figref{isopot} also furnishes an exactly 
solvable system~\cite{Kha89}, which makes them an interesting object for future studies in their own right. 
\begin{figure}[!htb]
\centerline{{\epsfxsize 14.5cm \epsffile[50 60 570 470]{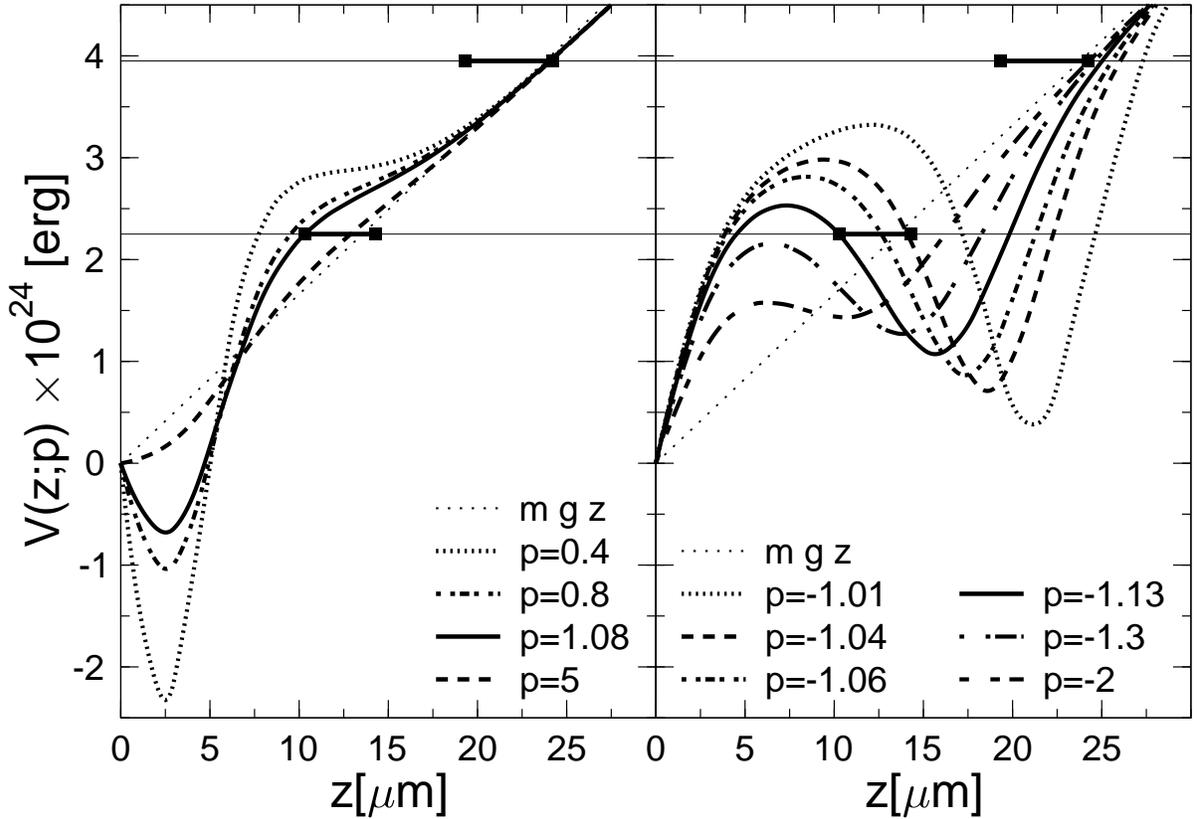}}}
\caption[f1]{Examples of isospectral potentials of a neutron, \eqref{isopot}, 
for $p > 0$ (left panel) and $p<-1$ (right panel).
The energies of the ground and first excited states $E_1 = 2.25 \times 10^{-24}$ erg and
$E_2 = 3.93 \times 10^{-24}$ erg are shown by the horizontal lines, and the diagonal dotted line is the Newtonian potential $m g z$. 
The experimental uncertainties of the measured ground state turning point $z_1^{exp} \in [10.3, 14.1]\, \mu m$
and of the excited state $z_2^{exp} \in [19.2, 24.0]\, \mu m$ are indicated by the intervals between the squares on the corresponding energy levels. The potentials compatible with the measured turning point  
cross the horizontal line within these interval.        
\figlab{isopot}}
\end{figure}

The ground state wave function corresponding to the isospectral potential~\eqref{isopot} can be written 
\beq
\psi_1(z;p) = C(p) \frac{\psi(z)}{{\mathcal{I}}(z) +p} \,,
\eqlab{isofun}
\eeq
as can be checked by substitution into the Schr\"{o}dinger equation. 
The proportionality constant $C(p)$ 
can be obtained from the normalization condition
$ \int_0^\infty \! dz \psi_1^2(z;p) = 1$:
\beq
\left| C(p) \right| = 
1 \bigg/ \sqrt{ \int_0^\infty \! d z \left[ {\psi_1(z) \over {\mathcal{I}}(z)+p} \right]^2}=\sqrt{p(p+1)} \,.
\eqlab{norm}
\eeq
Note that the wave function changes with changing $p$, while nevertheless corresponding to the same ground state energy level $E_1$.
The original potential and ground state wave function are restored from their isospectral counterparts in the limits $p \rightarrow \pm \infty$. 
 
All isospectral potentials $V(z;p)$ have the same full spectrum of energies as
the original Newtonian potential $V(z)$, including the ground state and the excited states. 
By contrast,~\eqref{isofun} is valid for the ground state wave function $\psi_1(z;p) \equiv \psi(z;p)$ only.
The normalized wave functions $\psi_{n+1}(z;p)$ of the excited states of the isospectral hamiltonians can also be constructed~\cite{Kha89} from their unmodified counterparts $\psi_{n+1}(z)$ ($n=1,2,3,\ldots$):
\beq
\psi_{n+1}(z;p)=\psi_{n+1}(z) + \frac{\hbar^2}{2 m E_{n+1} \left[ {\mathcal{I}}(z)+p \right]} \, 
\frac{d {\mathcal{I}}(z)}{d z} 
\left[ \frac{d}{d z} - \frac{1}{\psi_1(z)} \frac{d \psi_1(z)}{d z} \right] \psi_{n+1}(z) \,,
\eqlab{ex_fun}
\eeq 
where the energies $E_{n+1}$ and the integral ${\mathcal{I}}(z)$ are given in 
Eqs.~(\ref{eq:energies}) and (\ref{eq:int}), respectively.

\section{Experimental constrains on the isospectral potentials} \seclab{constr}

In this section we 
will determine the values of $p$ for which the isospectral potentials 
$V(z;p)$ are compatible with the experimentally observed~\cite{Nes02,Nes05} neutron gravitational states. 
The lowest height at which the flux of ultra-cold neutrons has a ``step", corresponding to the lowest quantum state, has been measured to be  
$z_1^{exp} = (12.2 \pm 1.8_{syst} \pm 0.7_{stat}) \mu m$, with the estimated total error
$\Delta z_1^{exp} \approx 1.9 \, \mu m$.
The (much less pronounced) next leveling of the flux, corresponding to the first excited state, has been extracted to occur at
$z_2^{exp} = (21.2 \pm 2.2_{syst} \pm 0.7_{stat}) \mu m$, yielding a total error
$\Delta z_2^{exp} \approx 2.3 \, \mu m$. Within the error bars, these heights are consistent with the theoretical turning points
of the ground and first excited states associated with the 
Newton potential $m g z$ (e.~g., see Refs.~\cite{Flu74,Vor06}). 
It should be emphasized, however, that $z_2$ has 
so far been measured with much worse resolution than $z_1$~\cite{Nes05}; therefore, in our model, we will treat the constraints stemming from $z_2^{exp}$ as less firm than those from $z_1^{exp}$.  

Our objective is to select those isospectral potentials $V(z;p)$ for which 
the semiclassical turning points of a neutron with the energy $E_{1,2}$,
obtained from the equations $V(z_{1,2};p)=E_{1,2}$, lie within the
experimental error bars, i.~e.~$z_1^{exp} \in [10.3, 14.1]\, \mu m$
and $z_2^{exp} \in [10.2, 24.0]\, \mu m$.
It can be seen from~\figref{isopot} that the unmodified Newtonian potential $m g z$ itself 
satisfies this condition, as it should. 
Among the isospectral potentials, only those with $p \ge 1.08$ and $p \in [-1.13, -1.04]$ are compatible with the measured ground-state turning point $z_1^{exp}$, i.~e.~they cross the ground state energy level between the squares indicating the experimental uncertainty interval. The error bar associated with $z_2^{exp}$
does not put additional constraints on the potentials with $p > 0$, but for $p < -1$
it appears to marginally allow only the potential with $p = -1.13$. 
However, we should reiterate here that the constraints stemming from $z_2^{exp}$ are at present much less reliable than those from $z_1^{exp}$.
(The potentials with $p \alt -20$ also fulfill our selection criteria, but their deviations from the unmodified $m g z$ potential are extremely small; therefore, we do not show them and will not discuss them further.)

It is interesting to note that the ground state wave functions corresponding to the range
$p \ge 1.08$ have a behaviour qualitatively different from those corresponding to
$p \in [-1.13, -1.04]$. In particular, the latter wave functions describe neutrons predominantly localized in a potential well above the turning point, which is quite distinct from the wave function for the usual Newton potential, or the isospectral potentials with $p > 0$.
In the next section we will 
exploit the properties of the ground state wave functions to put additional phenomenological constraints on the potentials with $p > 1.08$ and $p \in [-1.13, -1.04]$, 
and in~\secref{excited} we will consider the contributions from the first excited state for the potentials with $p = 1.08$ and $p = -1.13$.

\section{Ground state neutron flux}\seclab{flux}

The experimental method~\cite{Nes02} is based on
a measurement of the step-like shape of the flux of the neutrons passing through a slit between the bottom mirror and the top absorber. The measured neutron flux is typically analyzed in detailed models including the interaction of the neutrons with the absorber, which is related to finite lifetimes of the quantum states.
Such analyzes were developed in Refs.~\cite{Vor06,Wes07}, where no modifications to Newton's potential were considered, while some assumptions were made about the widths of the quantum states, and
the neutron energies and populations of the states were fitted to 
reproduce the measured neutron flux as a function of the slit height. A careful modeling of the absorber is
crucial in these calculations to achieve a good description of the data.

By contrast, the focus of the present model is to explore the possibility that 
isospectral potentials could describe small-range deviations from the Newton gravitational potential.
In this section we study whether it might be possible to use experimental data to put additional constraints on the allowed isospectral potentials obtained above, 
all of which are compatible with the experimentally observed ground state as described above.
(The contributions of the excited states will be considered below in~\secref{excited}.)
To this end, for each of the isospectral hamiltonians, we will calculate the neutron flux,
assuming for simplicity that the states have zero widths and thus no interaction with the absorber is included. Due to these simplifying assumptions, a direct comparison with the existing experimental data is problematic, since an absorber is an important part of the actual experimental 
setup. Ideally, our
results should be compared with a neutron flux measured without an absorber.
Such an experiment would directly study properties of gravity itself,  
but a discussion of its feasibility lies beyond the scope of this paper. 
With this caveat, we will nevertheless proceed and use the data from Refs.~\cite{Nes02}, treating it as an approximate guide to study our calculated fluxes. 
 
As explained in~\secref{constr}, the ground state turning point $z_1^{exp} \in [10.3, 14.1]\, \mu m$ 
allows the potentials with $p \ge 1.08$ and
$p \in [-1.13, -1.04]$ (see~\figref{isopot}).      
\begin{figure}[!htb]
\centerline{{\epsfxsize 14.5cm \epsffile[50 60 570 470]{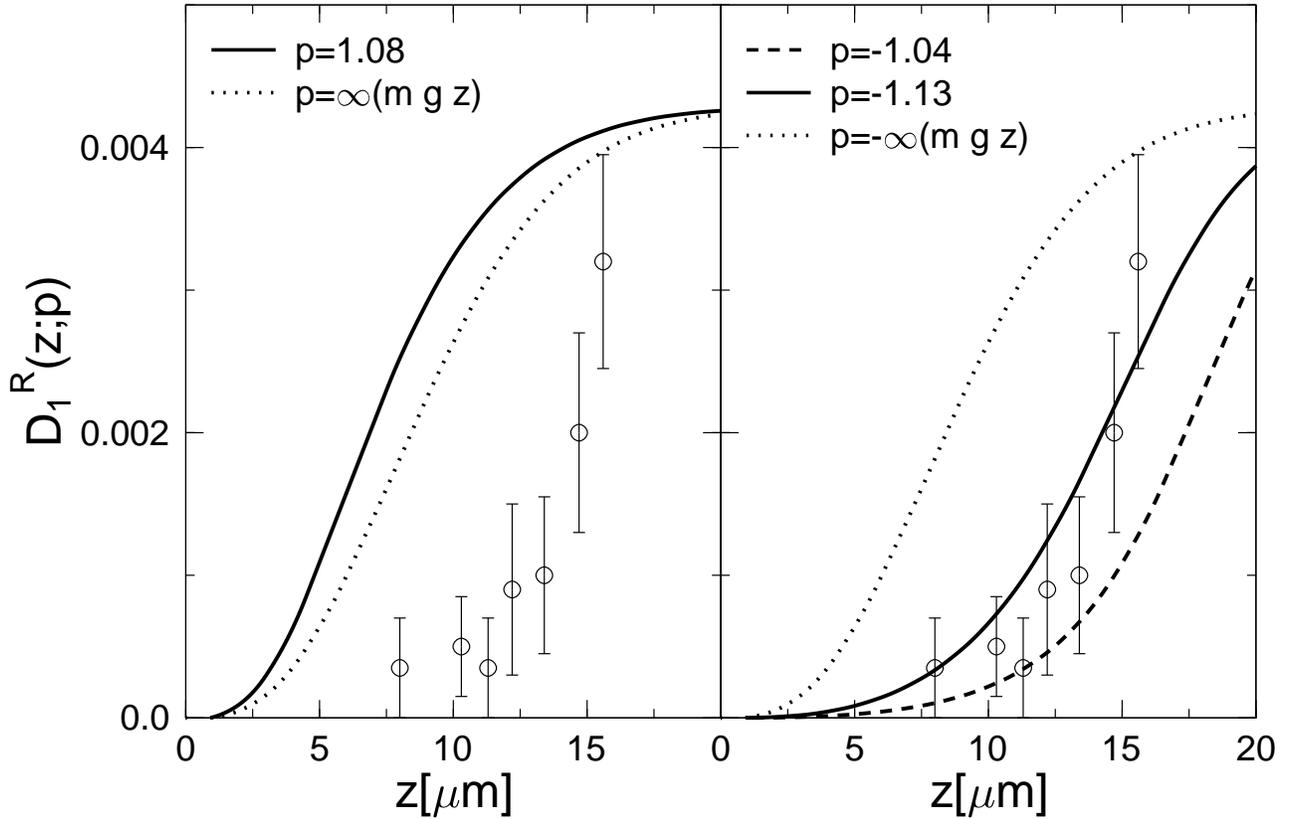}}}
\caption[f2]{The ground-state density integral~\eqref{dens}, renormalized as explained in the text to facilitate the comparison with the measured neutron count rate~\cite{Nes05}.  
The parameters $p$
limit the range of the isospectral potentials compatible with experimental data for the ground state, as described in~\figref{isopot}.  
\figlab{flux}}
\end{figure}
Representing the three-dimensional wave function of a neutron as a product of a plane wave along the horizontal direction and the one-dimensional ground-state wave function $\psi_1(z;p)$ along the vertical 
($z$) direction,
the counting rate of the neutrons in the detector is proportional to the average horizontal velocity of the neutrons and to the integral
\beq
D_1(z;p) = \int\limits_{0}^z \!d z' \psi_1^2(z';p)\,.
\eqlab{dens}
\eeq
We will call $D_1(z;p)$ the ground-state ``density integral". More precisely, it is the probability of finding a neutron in the ground state $\psi_1(z;p)$ above the mirror at any height less than or 
equal to $z$. An analogue of $D_1(z;p)$ is called
``the neutron flux" in Ref.~\cite{Vor06}. Since, apart from normalization factors, 
all these quantities have the same $z$-dependence as the measured neutron count rate, we will use the terms ``neutron flux", ``count rate" and ``density integral" interchangeably. 
 
We are interested in the behaviour of the density integral
for the parameter $p$ in the intervals compatible with the measured $z_1^{exp}$ as described above, 
i.~e.~ $p \ge 1.08$ and $p \in [-1.13, -1.04]$, and, for brevity, we will consider 
only the limiting values: $p={1.08,-1.04,-1.13}$, as well as the flux calculated for the standard $m g z$
potential (corresponding to $p \rightarrow \pm \infty$, as explained in~\secref{isospec}).    
The quantity plotted in~\figref{flux} is a renormalized density integral $D^R_1(z;p)=R_1 D_1(z;p)$, where
the constant $R_1$ does not change the $z$-dependence of the flux and is used for  
simply to facilitate the comparison with the data from Ref.~\cite{Nes05}. 
For completeness, we list the renormalization constants in the Appendix. 

Focusing on the behaviour of $D_1(z;p)$ near 
the lowest turning point $z_1^{exp} \in [10.3, 14.1]\, \mu m$, where the experimental flux exhibits a sharp 
step-like rise, 
we see that the rate of increase of the flux in the vicinity of $z_1^{exp}$ strongly depends on the value of parameter $p$. This suggests that, within our model, we can select the most likely isospectral potential by comparing the
corresponding density integral $D_1(z;p)$ with the shape of an experimental flux.
A fairly good agreement with the calculated flux is obtained for the isospectral hamiltonians with 
$-1.13 < p \alt -1.04$. It is noteworthy that,
apart from the trivial data renormalization, at this stage we do not have any adjustable 
parameter as the range of the $p$ was fixed independently in~\figref{isopot}.
The description of the data in~\figref{flux} is not perfect, due to the  
difficulties of a direct use of the existing data in our model, as explained in the beginning of this section. Note that $D_1(z;p=1.08)$ cannot satisfactorily describe the data, even with a variety of different normalizations applied. The flux for the usual 
$m g z$ potential also disagrees with the data; this disagreement emphasizes the importance of including an absorber in theoretical models based on the {\em unmodified} Newton potential, as was done in
Refs.~\cite{Vor06,Wes07}. 

We can see that the theoretical fluxes for $-1.13 < p \alt -1.04$ provide a good description of the onset of the flux ``step", which includes the data in the range of small $z$ up to the point of the sharp increase at 
$z \approx 13 \, \mu m$. 
The agreement deteriorates for $z \agt 16 \, \mu m$, 
where effects of excited states, not included in~\figref{flux}, become important. 
The excited states are described by 
the wave functions~\eqref{ex_fun}, and in principle, many of them can be included in the flux. However, since only the first excited state has been observed~\cite{Nes02,Nes05} with an acceptable resolution, we will study its contribution in some detail.

\section{Contribution of the excited state}\seclab{excited}

In this section we consider the isospectral potentials allowed by both the ground and first excited state, i.~e.~ those with $p=1.08$ and $p=-1.13$, as explained in~\secref{constr}. 
\begin{figure}[!htb]
\centerline{{\epsfxsize 16.0cm \epsffile[25 65 600 730]{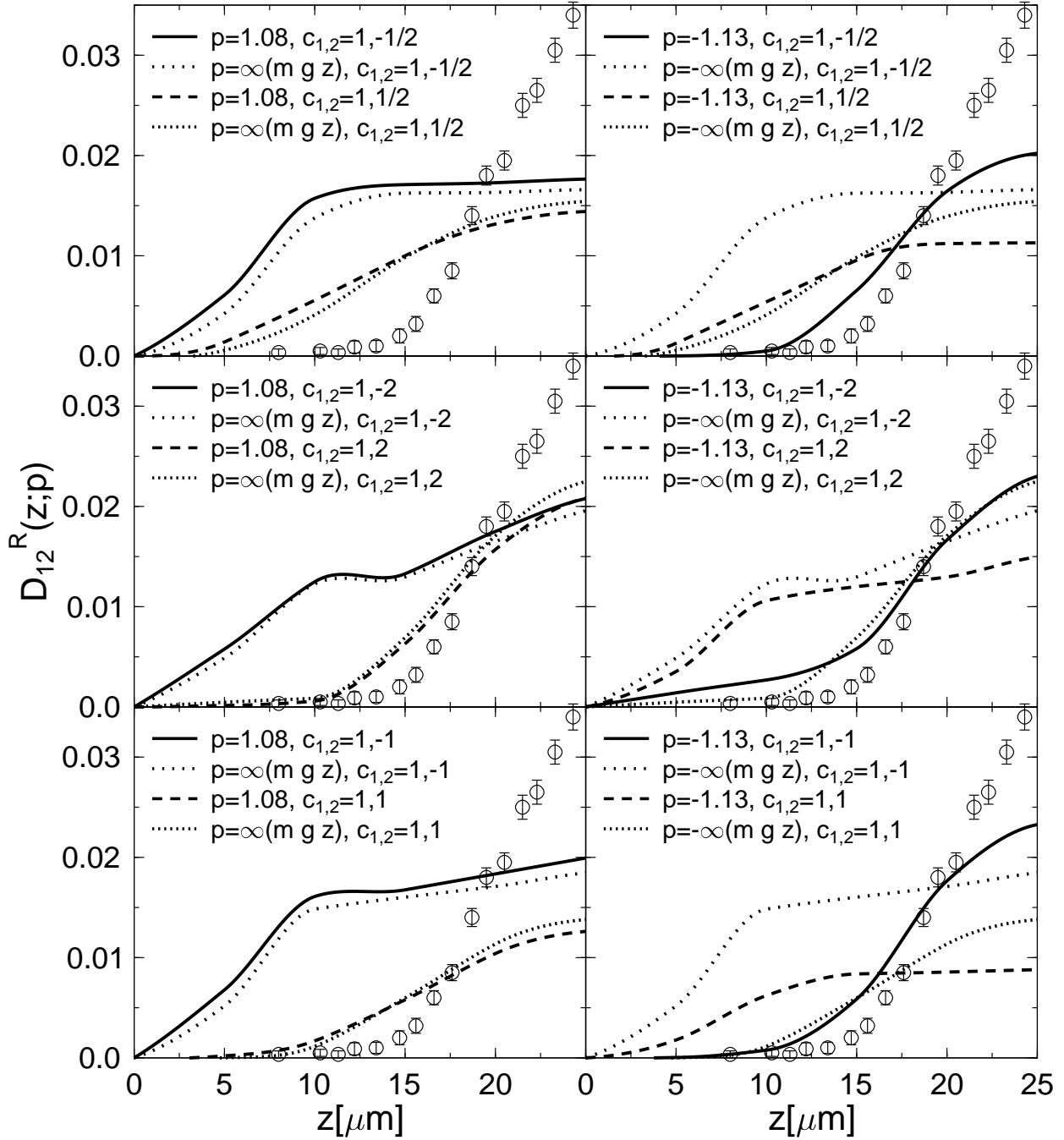}}}
\caption[f3]{The renormalized two-state density integral~\eqref{dens_mix} for the isospectral potentials with 
$p=1.08$ (left panels) and $p=-1.13$ (right panels), compatible with the measured turning 
points $z_{1,2}^{exp}$ as shown in~\figref{isopot}. The fluxes corresponding to several sets of constants
$c_{1,2}$ are shown from the top to bottom panels, to illustrate several distinct superpositions of the ground and first excited states.
The data points are from Ref.~\cite{Nes05}, and the flux renormalization factors are listed in Appendix. 
\figlab{flux_mix}}
\end{figure}
Assuming that a neutron is in a coherent superposition of the ground state $\psi_1(z;p)$, 
given by~\eqref{isofun}, and the
first excited state $\psi_2(z;p)$, given by~\eqref{ex_fun}, its wave function can be written as
\beq
\psi_{12}(z;p) = c_1 \psi_1(z;p) + c_2 \psi_2(z;p)\,, 
\eqlab{coh_wf}
\eeq
where $c_{1,2}$ coefficients describing the relative contributions of the states.
By analogy with~\eqref{dens}, we consider the ``two-state density integral" 
\beq
D_{12}(z;p) = \int\limits_{0}^z \!d z' \psi_{12}^2(z';p)\,, 
\eqlab{dens_mix}
\eeq 
which implicitly depends on $c_{1,2}$.

We will explore to what extent the two-state density 
integral $D_{12}(z;p)$ may be used to describe of the experimental flux up to the second 
turning point $z_2^{exp}$. 
Since the neutron populations in separate states were not directly 
measured~\cite{Nes05}, one can treat $c_1$ and $c_2$ as free parameters.
However, taking into account the simplicity of our model, we will not attempt to find the best quantitative fit to the data. Rather, our aim is to illustrate the contribution of the excited state for several contrasting choices of $c_1$ and $c_2$. 
Similar to the ground-state density integral $D_1(z;p)$, we renormalize $D_{12}(z;p)$ for comparison with the data and thus introduce $D_{12}^R(z;p)=R_{12} D_{12}(z;p)$. As explained in~\secref{flux}, this renormalization does not affect the important $z$-dependence of the flux. 
The method of fixing the renormalization constants 
$R_{12}$ and their values are described in the Appendix. 

\figref{flux_mix} shows the renormalized
two-state density integral $D_{12}^R(z;p)$ calculated for the isospectral potentials 
with $p=1.08$ and $p=-1.13$, as well as for the $m g z$ potential, and for several sets of $c_{1,2}$.
The bottom panels correspond to the cases where the excited state contributes with the same weight as the ground state, with two selections of signs: $c_1=\pm c_2$. The excited state contribution is emphasized in the middle panels ($|c_2/c_1|=2$), and attenuated in the bottom panels ($|c_2/c_1|=1/2$).
When $c_1$ and $c_2$ have opposite signs, the density integrals of the isospectral potential with $p=-1.13$ lie closer to the data than those of the potential with $p=1.08$ or the Newton 
potential $m g z$. For $c_1$ and $c_2$ of the same sign, the potential with $p=1.08$ or the $m g z$
perform better than that with $p=-1.13$. Among all calculations shown in~\figref{flux_mix}, the
case with $p=-1.13$ and $c_1=-c_2=1$ seems to describe the data better than the others.  
However, the agreement of the calculated fluxes with the data over the whole range of heights up to 
$z \approx 30 \, \mu m$ is rather poor. Some of the reasons for this discrepancy
have been already discussed in~\secref{flux}, such as the absence of an absorber in the present model.
On the other hand, an attractive feature of the model is that it uses only two parameters ($c_{1,2}$) to incorporate both the ground and the first excited state.
Such a simplified approach has the advantage of dealing exclusively with neutron properties under gravity, circumventing a possible explanation of the experimental 
results of~\cite{Nes02} as a ``neutron in a box" quantum system without any potential (see discussion in Refs.~\cite{Han03,Wes07}).
Assuming that the neutrons are subjected to gravity alone, without the influence of an absorber, 
Figs.~(\ref{fig:flux}) and  
(\ref{fig:flux_mix}) show that the fluxes calculated with the isospectral potentials can provide a qualitative description of the essential features of the data. 

Contributions of the higher excited states can be incorporated analogously by calculating a multi-state density integral 
$D_{12...N}$ (for $N$ states), with $N$ adjustable parameters $c_{1,2,...,N}$.
However, since the experimental resolution of
the higher excited states is at present quite poor~\cite{Nes05}, we will not extend the present analysis beyond the first excited state.

\section{Discussion and outlook} \seclab{disc}

There may be two possible interpretations of the results presented in this paper.
One possibility is to regarded this model as an effective approach which mimics models such as 
those of Refs.~\cite{Vor06,Wes07}, in which neutrons move under the usual Newton force, where a bottom mirror and a top absorber would be described in detail. Another possibility is to view the isospectral potentials considered here as true modifications of the Newton gravitational law at a micrometer scale. The characteristic length over which the deviation from the linear potential is observed is given by
$l = \left[ \hbar^2/(2 m^2 g) \right]^{1/3}$, which scales as $\sim m^{-2/3}$. This means that in experiments involving objects heavier than neutrons, modifications from the Newton law would be present on a shorter scale, which could explain why such modifications are
difficult to observe for atoms or macro-objects.    

It is interesting to note that in the vicinity of the turning point
$z_1 \approx 13 \mu m$ the isospectral potentials with 
$-1.13 < p \alt -1.04$ are alternately more attractive or more repulsive than the
unmodified Newton potential $m g z$. It is argued~\cite{Ark98,Bur04} that both the geometry of extra dimensions and the nature of quantum fields may influence whether the modification is attractive or repulsive. 

Our results suggest that the isospectral potential which most closely describe
the properties of the ground the first excited states, including the turning heights and (only qualitatively)the flux profile, is that with the parameter $p \approx -1.13$. 
A peculiar feature of this potential is a barrier of a certain width between $z=0$ and the lowest turning point $z_1 \approx 13 \mu m$ (see the right panel of~\figref{isopot}), which means that the small probability of finding a particle near $z \approx 0$ is a purely quantum effect of barrier penetration. Although such a gravitational potential looks rather unusual
from the classical mechanics point of view, it is perfectly legitimate since it has the same energy levels as the Newton potential and is also consistent with the 
experimentally observed properties of the neutron gravitational ground state.
As mentioned above, it would be interesting to
carry out an experiment similar to that of Refs.~\cite{Nes02,Nes05}, but measuring the
vertical distribution of the neutrons in a setup without an absorber. 
This would provide a more direct test of the present model.

One recently proposed~\cite{Bae09} 
direction of studying the excited gravitational states is to measure fluxes of 
spin-polarized ultra-cold neutrons in the presence of magnetic and gravitational fields, using the high-precision GRANIT spectrometer~\cite{Kre09}. 
Another improvement can be due to the use of new sources of ultra-cold neutrons,
notably the currently developed UCN source at TRIUMF~\cite{Mar09}, based on the earlier spallation design described in Ref.~\cite{Mas02}.
This and other high-density sources are expected to yield data with smaller error bars and a better resolution of the states.

\begin{acknowledgments}
We would like to thank Jeff Martin for useful comments and suggestions.
\end{acknowledgments}

\appendix
\section{Renormalization constants for the density integrals}\seclab{renorm}
\begin{table}[!hb]
\caption[t1]{Renormalization constants for comparing the density integrals with data.}
\begin{center}
\begin{tabular}{|c|c|}
\hline
 Renormalization factor & ``Best fit" flux  \\
\hline
$R_1 = 0.0043$ & $p=-1.13$   \\
\hline
$R_{12}=0.0071$ ($c_1=1, c_2=1)$ & $p=\pm \infty (m g z)$   \\
\hline
$R_{12}=0.0048$ ($c_1=1, c_2=2)$ & $p=1.08$   \\
\hline
$R_{12}=0.0125$ ($c_1=1, c_2=1/2)$ & $p=\pm \infty (m g z)$   \\
\hline
$R_{12}=0.0096$ ($c_1=1, c_2=-1)$ & $p=-1.13$   \\
\hline
$R_{12}=0.0042$ ($c_1=1, c_2=-2)$ & $p=-1.13$   \\
\hline
$R_{12}=0.0135$ ($c_1=1, c_2=-1/2)$ & $p=-1.13$   \\       
\hline
\end{tabular}
\end{center}
\tablab{renorm}
\end{table}
In~\tabref{renorm} we summarize the constants $R_1$ and $R_{12}$ which are used to renormalize the density integrals Eqs.~(\ref{eq:dens}) and (\ref{eq:dens_mix}) for comparison with the data in 
Figs.~(\ref{fig:flux}) and (\ref{fig:flux_mix}), respectively. 
In practice, the renormalization constants were chosen so that, for each combination of $c_{1,2}$, one of the calculated fluxes described the data in the best achievable way; these ``best fits" are listed in the right column. After that, the same values of 
$R_1$ and $R_{12}$ were used for all values of the parameter $p$. It should be stressed that we did not
attempt to do a quantitative fit to the data over the whole range of allowed $p$ and $c_{1,2}$, which is justified within the present simplified model.


\end{document}